\documentclass[preprint]{aastex6}
\usepackage{amsmath}
\usepackage{graphicx}
\usepackage{natbib}

\DeclareMathOperator{\sech}{sech}

\begin{document}

\newcommand{\curl}{\nabla \times}
\newcommand{\grad}{\nabla}
\newcommand{\pardiv}[3]{\frac{\partial^{#3} {#1}}{\partial {#2}^{#3}}}
\newcommand{\adiv}[1]{\frac{D {#1}}{Dt}}
\newcommand{\adivlong}[1]{\frac{\partial{#1}}{\partial t} + (\mathbf{u} \cdot \nabla){#1}}
\newcommand{\BLQ}{BLQ}

%attempt to automate filling in model values in paper

\newcommand{\gmodel}{.00928}
\newcommand{\Lundquistmodel}{10^4}
\newcommand{\kmodel}{.1315}
\newcommand{\posmodel}{[86,18]}

	\title{Modeling a propagating sawtooth flare ribbon structure as a tearing mode in the presence of velocity shear}

\author{Jacob Parker and Dana Longcope}
\affil{Department of Physics, Montana State University,
  Bozeman, Montana 59717}


\turnoffedit

%\linenumbers

\begin{abstract}
	On April 18, 2014 (SOL2014-04-18T13:03) an M-class flare was observed by IRIS.  The associated flare ribbon contained a quasi-periodic sawtooth pattern that was observed to propagate along the ribbon, perpendicular to the IRIS spectral slit, with a phase velocity of $\sim 15$ km s$^{-1}$ \citep{Brannon2015}.  This motion resulted in periodicities in both intensity and Doppler velocity along the slit.  These periodicities were reported by \citet{Brannon2015} to be approximately $\pm0.5$ arcseconds in position and $\pm20$ km s$^{-1}$ in velocity and were measured to be $\sim180^{\circ}$ out of phase with one another. This quasi-periodic behavior has been attributed by others to bursty or patchy reconnection \citep{Brosius2015,Brosius2016} and slipping occurring during three-dimensional magnetic reconnection \citep{Li2015,Li2016}.  While able to account for periodicities in both intensity and Doppler velocity these suggestions do not explicitly account for the phase velocity of the entire sawtooth structure, or for the relative phasing of the oscillations. Here we propose that the observations can be explained by a tearing mode instability occurring at a current sheet across which there is also a velocity shear. \edit3{ Using a linear model of this instability we reproduce}\edit1{ the relative phase of the oscillations, as well as the phase velocity of the sawtooth structure.}   We suggest a geometry and local plasma parameters for the April 18 flare which would support our hypothesis.  Under this proposal the \edit1{combined spectral and spatial} IRIS observations of this flare may provide the most compelling evidence to date of a tearing mode occurring in the solar magnetic field.
\end{abstract}

%\date{Draft: \today}

\section{Introduction}

Chromospheric flare ribbons, observed in many flares, are believed to provide indirect evidence of magnetic reconnection occurring in the corona above.
The most widely cited interpretation of these elongated ribbons of chromospheric emission is the CSHKP model
\citep{Carmichael1964,Sturrock1968,Hirayama1974,Kopp1976}.  It holds that an eruption has opened magnetic field lines in regions of opposite sense, and the resulting regions are temporarily separated by a current sheet (CS).  Reconnection at this CS creates closed field lines which subsequently retract downward to form the post-flare arcade.  The reconnection and retraction energizes those field lines which have been reconnected, and that energy is deposited into the chromosphere to produce the observed emission.  The elongated ribbon-like structure thereby maps out, in the chromosphere, the locus in the corona at which reconnection is occurring; it is an image of the CS.

The structure and motions of flare ribbons have provided great insight into the manner in which magnetic reconnection occurs in solar flares.  Spreading motion perpendicular to the CS has been used to infer the progress of reconnection and to measure the reconnection electric field \citep{Forbes1984,Poletto1986,Qiu2002,Qiu2004,Isobe2005}.  Other motions, either parallel to the ribbon, or less orderly, provide evidence of reconnection more complicated than in simple two-dimensional models \citep{Warren2001,Fletcher2004,Qiu2009,Li2015}.  The detailed, fine structure of the flare ribbons has also been used as evidence of complex structure within the CS itself \citep{Nishizuka2009}.

Recent UV spectral and imaging observations of flare ribbons at extremely high spatial resolution, made by IRIS \citep{dePontieu2014}, have revealed distinctive quasi-periodic, sawtooth patterns in certain flare ribbons \citep{Li2015,Brannon2015,Brosius2015,Brosius2016}.  These may offer a new clue to some facet of magnetic reconnection.  At the moment, however, there are several competing hypotheses about the cause of these small-scale
sawtooth patterns in the flare ribbon.  The pattern was first reported by \citet{Li2015} based on IRIS observations of the ribbons of the 2014 Sep 10 flare (SOL2014-09-10T17:45).  They interpreted it as a signature of quasi-periodic slipping, generated by a mode of three-dimensional reconnection \citep{Aulanier2006}.  A similar pattern was observed by \citet{Brannon2015} and \citet{Brosius2015} in the ribbons of the 2014 April 18 flare (SOL2014-04-18T13:03).  \citet{Brosius2015} and subsequently \citet{Brosius2016} noted that individual \replaced{points on the ribbon}{pixels on the IRIS slit} exhibited quasi-period brightenings and Doppler shifts at $\sim 3$ minute intervals in a number of spectral lines observed by IRIS as well as {\em Hinode} EIS.  They attributed this quasi-periodic behavior to a bursty or patchy mode of magnetic reconnection.

\begin{figure}[htb]
\centerline{\includegraphics[width=6.5in]{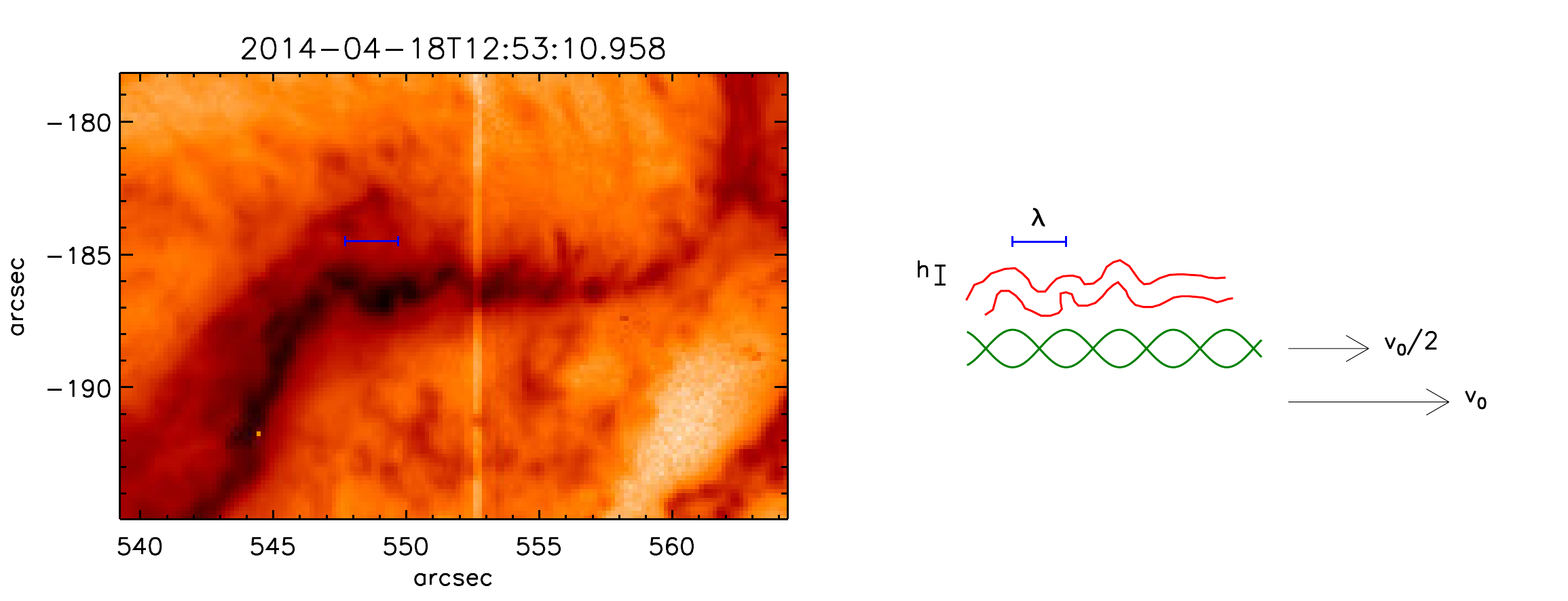}}
\caption{A frame from the IRIS 1400\AA\ SJI showing the flare ribbon from the 2014 Apr 18 flare.  A subsection of the full SJI is shown on the left using inverse log-scale, with axes marked in arcseconds from disk center.  The dark horizontal band is the flare ribbon, and a lighter vertical line at $x=552.7$ arcseconds is IRIS's spectrographic slit.  An outline of the ribbon is reproduced  in red on the right, along with notations showing the spatial period, $\lambda$, amplitude, $h$, and pattern velocity, $v_0/2$.  A green island-chain shows the proposed orientation of the tearing mode instability, displaced southward for clarity by a distance greater than proposed in the model.  A second velocity arrow shows the direction of the proposed shear velocity, $v_0$, which causes the island chain to move at $v_0/2$.}
	\label{fig:SJI}
\end{figure}

\citet[][hereafter called \BLQ]{Brannon2015} studied the same flare as \citet{Brosius2015} but reached a different conclusion about the cause of the quasi-periodic phenomena.  They studied both the 1394\AA\ and 1403\AA\ spectral lines of Si {\sc iv} ($T\simeq80,000$ K), over a range of positions along the spectral slit, as well as the sequence of 1400\AA\ slit-jaw images (SJIs), showing mostly the same Si emission.  The SJIs clearly revealed the sawtooth pattern in the flare ribbon with a spatial period of approximately $\lambda_{\rm st}=1.8$ Mm (roughly 2 arc seconds) shown here in Fig.\ \ref{fig:SJI}.  They also reported that the pattern seemed to propagate {\em along the ribbon} with apparent speed of $v_{\rm st}\simeq 15\,{\rm km\, s^{-1}}$ (denoted in Fig.\ \ref{fig:SJI} as $v_0/2$).  This parallel pattern speed was super-imposed on the much slower, $1$--$2\,{\rm km\, s^{-1}}$, outward motion of the ribbon itself (southward in Figure \ref{fig:SJI}).  As the pattern moved across the slit, it caused the point of peak intensity to move up and down at $\sim20\,{\rm km\, s^{-1}}$, with the $\sim 3$ minute period.   This apparent motion of the bright ribbon along the slit resulted in the quasi-periodic brightening {\em of a fixed \edit1{\textit{pixel}} position} reported by \citet{Brosius2015}.

% To be mentioned viz. Brosius: energization of a single, stationary footpoint is known to result in a sudden brightening, and is often observed in UV 
% lines including C iv from AIA 1600.  In such cases, however, the sudden brightening is followed by a far more gradual fading.  This is explained by the
% slow cooling of the coronal loop creating a high coronal pressure which results in enhanced EM at the TR.
%  The light-curves in Brosius do not show this gradual decay.  That suggests they are not from a single footpoint.

\begin{figure}[htb]
\centerline{\includegraphics[height=6.0in]{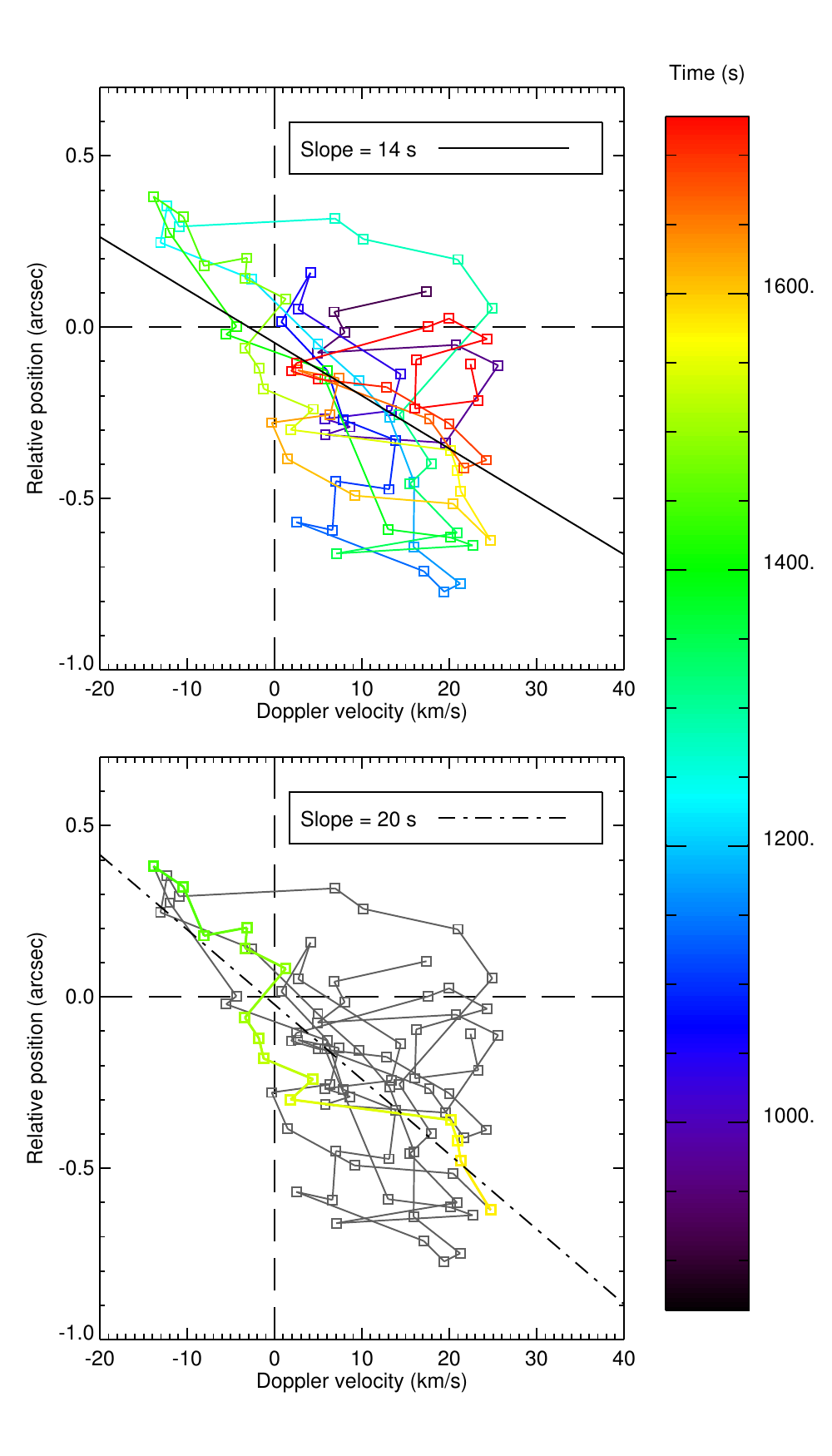}}
\caption{Figure 11 from \BLQ\ reproduced with permission.  The upper panel shows the mean Doppler velocity of the blue component of the Si {\sc iv} lines, plotted against the mean position of the sawtooth pattern along the slit.  Time is indicated by color, coded in the bar along the right.  The bottom panel repeats this, but highlights in color one of the best defined oscillations.}
	\label{fig:Brannon11}
\end{figure}

\BLQ\ successfully fit  each Si {\sc iv} spectral line with two Gaussians in about about $80\%$ of the ribbon's pixels.  The central positions of each Gaussian component oscillated in time, redward and then blueward. The bluer of the two components oscillated about a small mean velocity, and therefore showed red shifts followed by blue shifts of 
$\pm 20\,{\rm km\, s^{-1}}$. \edit1{The measurement of this secondary bluer component, as well as the phase velocity of the sawtooth feature, is the fundamental difference between the interpretation in this work and previous studies. \citet{Brosius2015} calculated Doppler velocities by finding the centroid of various line profiles, a procedure that might mask the Doppler velocities of the separate components.} The redder component oscillated about central value $\sim 40\,{\rm km\, s^{-1}}$, presumably from explosive chromospheric evaporation \citep{Fisher1985b}.  In both cases the velocity shift was approximately $180^{\circ}$  {\em out of phase} with the position shift: it was bluest when the pattern had reached its northern most \replaced{slit position}{point on the slit}.  This phase relation is depicted in Fig.\ 11 of \BLQ, which we reproduce here in Fig. \ref{fig:Brannon11}.  It is especially clear in a single oscillation, highlighted in the bottom panel, that the velocity {\em vs.} position curve forms a diagonal line ($180^{\circ}$ phase difference) rather than an ellipse ($90^{\circ}$ phase difference).

Since the Apr 18 flare occurred some $40^{\circ}$ from disk center, plasma motion which is horizontal at the solar surface would produce Doppler shifts as well as motion along the slit (\BLQ).  Kinematics of linear motion dictate that position and velocity oscillate $90^{\circ}$ out of phase from one another.  To explain the observed phasing \BLQ\ proposed that the plasma was moving in {\em horizontal ellipses}, so its two orthogonal components were $\sim90^{\circ}$ out of phase with one another.  In that case the velocity of one component (the line-of-sight) would be $\sim180^{\circ}$ out of phase with the {\em position} of the other (position along the slit), in agreement with the observations.  

\BLQ\ went on to observe that elliptical fluid motions occur in a number of well-known, large-scale instabilities occurring at some kind of surface or interface.  The classic Kelvin-Helmholtz instability \citep[KH,][]{Chandrasekhar1961} and the tearing mode \citep[{T}M,][]{Furth1963}, both occur at interface surfaces (a velocity shear layer and a CS respectively) and therefore have eigenmodes which are periodic along the interface and decay exponentially away from it.  Such a structure in the velocity stream function naturally produces elliptical fluid motions.  Indeed, a simple surface gravity wave (a.k.a.\ an $f$-mode) is well-known for its elliptical motions.  \BLQ\ therefore proposed that the quasi-periodic oscillation they observed in the flare ribbon was a manifestation of some form of surface-confined instability such as the KH or {T}M.

Both the KH and {T}M instabilities have been the subjects of extensive study, and have been previously invoked in numerous roles in solar phenomena. Si {\sc iv} line profiles with bright cores and broad wings have been associated with TM island formation, ``plasmoids'', during small scale reconnection and transition region explosive events \citep{Innes2015}.  Plasma blobs imaged by AIA \citep{AIA_instrument} have been observed to occur in the current sheet above a flare arcade \citep{Takasao2012,Takasao2016} and in coronal bright points \citep{Zhang2016}.  These plasma blobs are thought to be the result of magnetic islands generated by the TM instability. Wave-like structures observed on the edge of erupting coronal structures \citep{Ofman2011,Foullon2011,Foullon2013} with various phase velocities have been attributed to the KH instability. Most attributions to the TM and KH instability rely on appearance alone.  While we trust that those attributions are correct, we hope to show that the \edit1{spatial and spectral} IRIS observations of April 18 provide a more quantitative link between observation and model.

%It seems that with higher and higher resolution instruments the imprints of these instabilities are seen more and more frequently.

Either instability, in its distinct, traditionally-studied form, faces significant challenges in explaining the quasi-periodic pattern observed in the Apr 18 flare ribbon (\BLQ).  For shear in a magnetized plasma to undergo KH instability the velocity difference across the shear layer must exceed the component of Alfv\'en speed \replaced{tangent to the layer}{parallel to the flow difference} \citep{Chandrasekhar1961}.  The magnetic field strength near the ribbon was $\simeq 150$ G, leading to an Alv\'en speed $v_{\rm A}\sim 10,000\, {\rm km\, s^{-1}}$, some 500 times greater than any of the observed flow speeds.  While only a fraction of that field strength will be directed parallel to the flow, it seems extremely unlikely that the fraction could be as small as 0.2\%.  

Tearing modes, on the other hand, are well known features in CSs.  They will be unstable if the plasma resistivity is low enough that the Lundquist number, computed with respect to the thickness of the CS, exceeds $S\sim10^4$ \citep{Biskamp1986}.  Classical resistivity and a Sweet-Parker CS result in values as large as $S\sim10^8$.  The instability will produce elliptical motions, but at speeds lower than the Alfv\'en speed by an inverse fractional power of $S$.  Depending on the value of $S$, and on the value of the inverse power (this depends on other factors as well), that might be well below the $20\,{\rm km\, s^{-1}}$ flow speeds observed.  Furthermore, the traditional {T}M analysis, set in an otherwise stationary CS, does not predict a phase velocity with which we might associate the observed pattern speed of $v_{\rm st}\simeq15\,{\rm km\,s^{-1}}$.  

In the present work we show that when a steady \edit1{sub-Alfv\'enic} velocity shear occurs across a CS the modified {T}M instability can achieve higher flow speeds {\em and} exhibit a significant phase velocity.  Indeed, the two speeds will generally be comparable, as they were observed to be in the Apr 18 flare.  We find that such a combined KH/{T}M instability is compatible in many respects with the observations reported by \BLQ.  We therefore propose that the flare of 2014 Apr 18 provides evidence of a {T}M-like instability occurring at its CS.

While many previous studies of the {T}M have been set at a static CS \citep{Furth1963,Steinolfson1983,Velli1989,Pucci2014}, several notable exceptions did include an equilibrium shear flow
\citep{Einaudi1986,Ofman1991,Offman1993,Chen1997,Li2010,Zhang2011,Doss2015}.  Those analyses focused their attentions on effects the shear had on the growth rates of the instabilities.  We repeat these previous analyses here in order to compute the amplitude and structure of the eigenmode's velocity field. \edit1{A simple linear model allows us to easily generate flow structures while minimizing free parameters thereby permitting us to explore a wide swath of parameter space.}  From this we are able to synthesize the position and Doppler signature of a plasma element.  We find these reproduce reasonably well the aspects of the Apr 18 observation described above.  We propose that there was a horizontal bulk plasma motion of $v_0\simeq37\,{\rm km\,s^{-1}}$ on the open-field (unreconnected) side of the flare ribbon and no flow on the other (post-reconnection) side, as indicated in Fig.\ \ref{fig:SJI}.  In this configuration any tearing mode in the CS will propagate with a phase speed of $v_0/2\simeq 18.5\,{\rm km\,s^{-1}}$\edit1{, the average of the far field equilibrium shear flows,} as observed \edit1{\citep{Doss2015}}.  The fluid elements on the stationary side of the CS will then undergo elliptical motion, with similar velocity, about a fixed central point.  Since this is the post-reconnection side, it harbors the energized ribbon and is the plasma whose emission we observe.  This, we believe, was what occurred and was observed by IRIS on 2014 Apr 18.  If so, it provides a novel observational characterization of a tearing mode and the CS hosting it.  It may offer the most compelling evidence to date of a tearing mode occurring in the solar magnetic field.

We present our analysis and support our conclusion as follows.  In the next section we reprise the linear eigenmode calculation  for tearing modes at a CS at which there is also an equilibrium  shear flow \citep{Einaudi1986,Ofman1991}; they are a form of combined TM/KH mode.   We forego discussion of growth rates and focus instead on the structure of the velocity fields of the eigenmodes \edit1{which were not thoroughly presented by previous investigators}.  Then in Sec.\ 3 we propose a geometry in which such an instability might have occurred in the 2014 Apr 18 flare.  We synthesize the observations which would result from such a scenario.  The synthesized observation compares favorably to the observations reported by \BLQ, and to Fig.\ \ref{fig:Brannon11} in particular.  Finally, in Sec.\ 4 we discuss the significance of our proposed explanation.

\section{Linear Tearing Mode Instability with Shear Flow}

	Since the first publication on the linear tearing mode instability in resistive plasmas \citep[FKR]{Furth1963} many investigators have found solutions to the eigenvalue problem numerically.  These numerical studies tested the validity of assumptions used to find the instability growth rates and scaling relations analytically \citep{Steinolfson1983}. They also examined the effects of extra terms in the governing equations, different coordinate systems, and boundary conditions on instability growth \citep{Velli1989,Ofman1991}. We repeat this analysis in order to verify the observational effects the TM instability with an asymmetric shear flow.  The following solution method has been used in many of those previous investigations \citep{Killeen1970,Steinolfson1983,Ofman1991} and is included here for completeness.
			
	For our analysis we use the two-dimensional, incompressible, resistive magnetohydrodynamic (MHD) equations.  The TM instability is governed by the resistive induction equation,  
	\begin{equation}
		\frac{\partial\mathbf{B}}{\partial t}  = \curl (\mathbf{u} \times \mathbf{B}) - \curl (\eta \curl \mathbf{B})~~,
	\end{equation}	
vorticity equation,
	
	\begin{equation}
		\frac{\partial (\curl \mathbf{u})}{\partial t}
		 = \curl\Bigl\lbrace \frac{1}{4 \pi \rho_0} (\curl \mathbf{B}) \times \mathbf{B} - (\curl \mathbf{u}) \times \mathbf{u} \Bigr\rbrace~~, 
	\end{equation} 
and condition of incompressibility, 
	\begin{equation}
		\nabla \cdot \mathbf{u} = 0~~.
	\end{equation}
We also take resistivity, $\eta$, and density, $\rho_0$, to be uniform.
	
	We linearize these equations about an initial magnetic field, $\mathbf{B}_0(y)$, given by the Harris current sheet \citep{Harris1962}, along with an initial velocity shear profile $\mathbf{u}_0(y)$.  For simplicity we give the shear profile the same functional form as the magnetic field. In Section 3 we consider an asymmetric velocity shear, however, to simplify analysis in this section we move to a reference frame in which the velocity shear is symmetric: the flow is equal and opposite across the current sheet. In this reference frame, for sufficiently small shear velocity $V$, the phase velocity of any instability will be zero.  The instability will then be given a phase velocity when, in Sec. 3, we return to the stationary frame of reference.  For further simplicity, the thickness, $a$, of the current sheet and velocity shear are taken to to be the same
		\begin{equation}
			\mathbf{B}_0 (y)= B_0 \tanh(y/a) \ \bf\hat{x}~~,
		\end{equation}
		\begin{equation}
			\mathbf{u}_0(y) = V \tanh(y/a) \ \bf\hat{x}~~,
		\end{equation}
where $B_0$ and $V$ are the far-field magnitudes of the field and velocity respectively.
		
	We perturb all quantities with first order terms of the form $ f_1(x,y,t) = f_1(y,t) \ e^{ikx}$. Using incompressibility to eliminate the $\bf\hat{x}$ component of $\mathbf{u}_1$, the solenoidal magnetic field condition to eliminate the $\bf\hat{x}$ component of $\mathbf{B}_1$,  and taking only the $\bf\hat{z}$ component of the vorticity equation gives a coupled set of second-order linear differential equations. We non-dimensionalize the equations by scaling magnetic field to the asymptotic field strength, $B_0$, velocity to the asymptotic Alfv\`{e}n speed, $v_a=B_0/\sqrt{4\pi\rho_0}$ , and lengths to the thickness of the current sheet, $a$.  The resulting non-dimensional quantities are denoted with over-bars
		\begin{align}
		\label{eqn:nodim} 
			\bar{u} &= u_{1y}/v_a,  &\bar{B} &= B_{1y}/B_0,  &\bar{t} &= \frac{tv_a}{a},  \\
			\bar{k} &= ka,  & S &= \frac{a v_a}{\eta}, & \bar{y} &= y/a~~.
		\end{align}
% \bar{\gamma} &= \gamma v_a/a  --- this quantity does not appear, so I have removed it from the defs.
%  ALSO - there is an error in the defintion
	 The non-dimensionalized profiles are denoted
			\begin{equation}
		F(\bar{y}) = \frac{B_{0x}}{B_0} = \tanh(\bar{y}), \quad  G(\bar{y})= \frac{u_{0x}}{v_a}=\bar{V}\tanh(\bar{y}),
		\end{equation}
In terms of these, the non-dimensional linearized equations are,
	\begin{equation}
		\label{eqn:tm_in}
			\pardiv{ \bar{B}}{\bar{t}}{} = i\bar{k}(F\bar{u}-G\bar{B}) + S^{-1}\mathcal{L}\bar{B}
		\end{equation}
		and
		\begin{equation}
		\label{eqn:tm_momentum}
			\pardiv{ \bar{u}}{\bar{t}}{} = i\bar{k}\lbrace F\bar{B}-G\bar{u} + \mathcal{L}^{-1}(G''\bar{u}-F''\bar{B})  \rbrace
		\end{equation}
where primes denote derivatives with respect to $\bar{y} \equiv y/a$, and we have introduced the linear operator
		\begin{equation}
		\label{eqn:L}
			\mathcal{L} = \pardiv{}{\bar{y}}{2} - \bar{k}^2,
		\end{equation}
		whose inverse is denoted	$\mathcal{L}^{-1}$. 
		
Our velocity variable, $\bar{u}$, differs from other presentations of TM eigenfunctions \citep{Velli1989,Ofman1991} by a factor of $i$.  Their definition is preferable in the absence of shear flow because our version of $\bar{u}$ becomes purely imaginary in that case.  When shear is present ($V\ne0$) we find the more symmetric definition of $\bar{B}$ and $\bar{u}$, given in Equation (\ref{eqn:nodim}), to be preferable. 
	
	Considering separately the real and imaginary parts of Equations (\ref{eqn:tm_in}) and (\ref{eqn:tm_momentum}) gives four equations to advance four independent fields: the real and imaginary parts of $\bar{u}$ and $\bar{B}$. We solve these equations numerically by implicitly integrating all four components forward in time until the fastest growing unstable mode dominates and pure exponential growth sets in at a single, real growth rate $\bar{\gamma}=\gamma a/v_a$. At every time step the eigenfunctions are re-scaled to make the largest value of the largest field (typically the real part of $\bar{B}$) unity. The growth rate for that time step is calculated to be
	\begin{equation}
	\bar{\gamma} = \frac{\ln(\max([\bar{B},\bar{u}]))}{\Delta \bar{t}},
\end{equation}	
prior to the solutions being re-scaled.	
	 
	 The solution is said to have converged on the fastest growing mode when
	\begin{equation}
	\frac{\sum_i |\bar{u}_{i,j}-\bar{u}_{i,j-1}|^2}{\sum_i |\bar{u}_{i,j-1}|^2} < 1 \times 10^{-9},
	\end{equation}
where $i$ is a spatial index and $j$ is the time step. The velocity eigenfunction, $\bar{u}$, is used to test convergence because it is found to change more over the course of simulation and therefore takes longer to converge. At the point of convergence the last calculated value of $\bar{\gamma}$ is the growth rate for that mode.
	
		A purely real $\bar{\gamma}$ and odd equilibrium functions $F$ and $G$ mean that the real parts of $\bar{B}$ and $\bar{u}$ will be even in $\bar{y}$ while the imaginary parts will be odd,
	\begin{equation}
	\mathrm{Re}[\bar{u}(y)] = \mathrm{Re}[\bar{u}(-y)], \quad \mathrm{Im}[\bar{u}(-y)] = -\mathrm{Im}[\bar{u}(y)],
	\end{equation}
	\begin{equation}
	 \mathrm{Re}[\bar{B}(y)] = \mathrm{Re}[\bar{B}(-y)], \quad \mathrm{Im}[\bar{B}(-y)] = -\mathrm{Im}[\bar{B}(y)].
	\end{equation}
	To enforce these symmetries we set to zero the values of the imaginary parts and derivatives of the real parts of both $\bar{u}$ and $\bar{B}$ at $\bar{y}=0$. This permits us to solve in the half space $\bar{y}  \geq 0$.  Any normal modes with a complex growth rate (i.e. non-vanishing phase velocity) will not have these symmetries.  In this case all four components of the solution will have both odd and even parts and a solution in full space would be necessary. We do not attempt this in the current study. 
	  
	  Spatial derivatives are computed using a second-order finite differencing scheme on a nonuniform grid. We use a grid from \citet{Steinolfson1983} with spacing prescribed as
	  \begin{equation}
	   \Delta \bar{y}_i=\bar{y}_{i+1}-\bar{y}_i = \Delta\bar{y}_{\max}\Bigl\lbrace \frac{\Delta \bar{y}_{\min}}{\Delta\bar{y}_{\max}} \Bigr\rbrace^{(i_m-i)/(i_m-1)}.
      \end{equation}
For our purposes we used a minimum spacing near the origin,$\Delta \bar{y} = 1\times 10^{-4}$, that increases to a max spacing, $\Delta \bar{y} = .5$, at grid point $i_m = 300$.  Beyond this point a spacing of $\Delta \bar{y}_{/max}$ is maintained to the outer boundary $\bar{y}=1000$.
      	   	
Our outer boundary condition is set to match the solutions onto the asymptotic analytic solutions to Equations (\ref{eqn:tm_in}) and (\ref{eqn:tm_momentum}).  There are four asymptotic solutions to these equations, two exponentially growing and two exponentially decaying.	 One exponential decay solution has a decay rate that is proportional to $S$ and therefore decays much faster than the other for the large Lundquist number.  We assume that this mode is absent far from current sheet and impose the more slowly decaying asymptotic solution as a boundary condition. This gives

	\begin{equation}
	 \lim_{y \rightarrow \infty} \bar{u}(|y|) \propto e^{-k|y|},
	 \end{equation}
	 \begin{equation}
	\lim_{y \rightarrow \infty} \bar{B}(|y|) \propto e^{-k|y|}.
	\end{equation}
We implement this numerically by enforcing
	\begin{equation}
	f(\bar{y}_n) = f(\bar{y}_{n-1})e^{-\bar{k}(\bar{y}_n-\bar{y}_{n-1})}
	\end{equation}
at the outer most grid point, $n$, for both $\bar{u}$ and $\bar{B}$.

	To obtain a single solution we fix $S$, $\bar{k}$, and $V$ and integrate forward in time using a backward Euler method. The time evolution operator for this method does not have explicit time dependence and therefore only needs to be inverted once for every run.  Since we are not interested in the time dependent behavior of these solutions a low order integration scheme can be used without loss of accuracy. The stability that comes from integrating implicitly allows for large time steps, $\Delta \bar{t} \sim 100$, and leads to very fast convergence in roughly a hundred time steps.  The choice of time step and convergence time depend mostly on the choice of $S$. Larger time steps can be taken when the growth rate is slower, or for larger $S$.  
		
\begin{figure}
		
		\caption{The eigenfunction for the most unstable mode with parameters $S = 1 \times 10^7$, $V = 1.5 \times 10^{-3}$, and $\bar{k} = .0234$ plotted along a logarithmic spatial coordinate. This mode has a growth rate $\bar{\gamma}= 2 \times 10^{-4}$. }
		\label{fig:eigen}
		\centerline{\includegraphics[scale=.6]{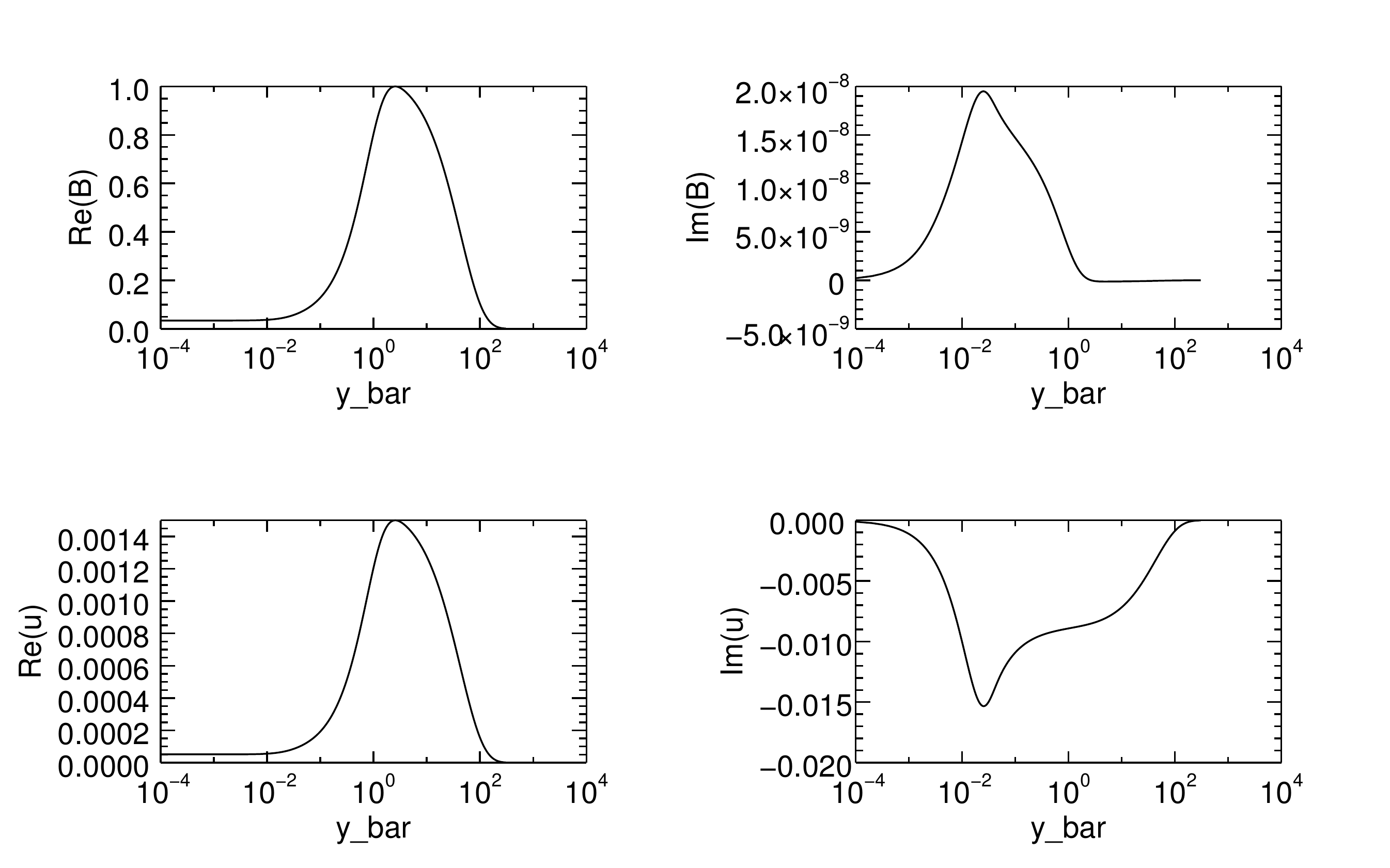}}
	\end{figure}

	Figure \ref{fig:eigen} shows an example of the eigenfunctions for the TM instability with an equilibrium shear flow $\bar{V} = 1.5 \times 10^{-3}$, $\bar{k}=\bar{k}_{\rm max}=0.0234$, and $S = 1\times 10^7$.  Here $\bar{k}_{\rm max}$ is the wavenumber leading to a maximum growth rate for the given value of $S$.  The magnetic eigenmode is fairly uniform 
($\bar{B}\simeq0.03$) out to about $\bar{y}\simeq0.03$.  The velocity eigenmode has structure outside this region, and thus does not conform the the so-called constant-$\psi$ approximation (FKR) frequently used to solve for eigenmodes analytically. This is consistent with the findings of \citet{Steinolfson1983} who found that the maximum growth rate for a given $S$ always occurs in an area of parameter space where this approximation is not valid.
	
	To visualize these eigenfunctions we contour the flux function, $A$, and stream function, $\Phi$, defined to produce the magnetic field and velocity fields
	\begin{equation}	
	 \mathbf{B} = \nabla \times A(x,y) \mathbf{\hat{z}}~~,
	 \end{equation}
\begin{equation}	
	 \mathbf{u} = \nabla \times \Phi(x,y)\mathbf{\hat{z}}~~.
	 \end{equation}
The full value of each consists of an equilibrium and a perturbed contribution
	 \begin{equation}
	 A_{0} = \ln(\sech(y)) \quad,\quad A_{1} = \frac{1}{\bar{k}}\lbrace \mathrm{Re}(\bar{B})\sin(\bar{k}\bar{x}) + \mathrm{Im}(\bar{B})\cos(\bar{k}\bar{x})\rbrace~~,
	\end{equation}

	 \begin{equation}
	 \Phi_{0} = \bar{V}\,\ln(\sech(y)) \quad,\quad \Phi_{1} = \frac{1}{\bar{k}}\lbrace \mathrm{Re}(\bar{u})\sin(\bar{k}\bar{x}) + \mathrm{Im}(\bar{u})\cos(\bar{k}\bar{x})\rbrace~~.
	\end{equation}

\begin{figure}
	\caption{Contours of $\bar{B}$ (black) and $\bar{u}$ (red) with shear flow $V$ = 0, 1.5 $\times 10^{-3}$, 0.1 in a,b,c respectively.}
		\label{fig:contour}
			\gridline{\fig{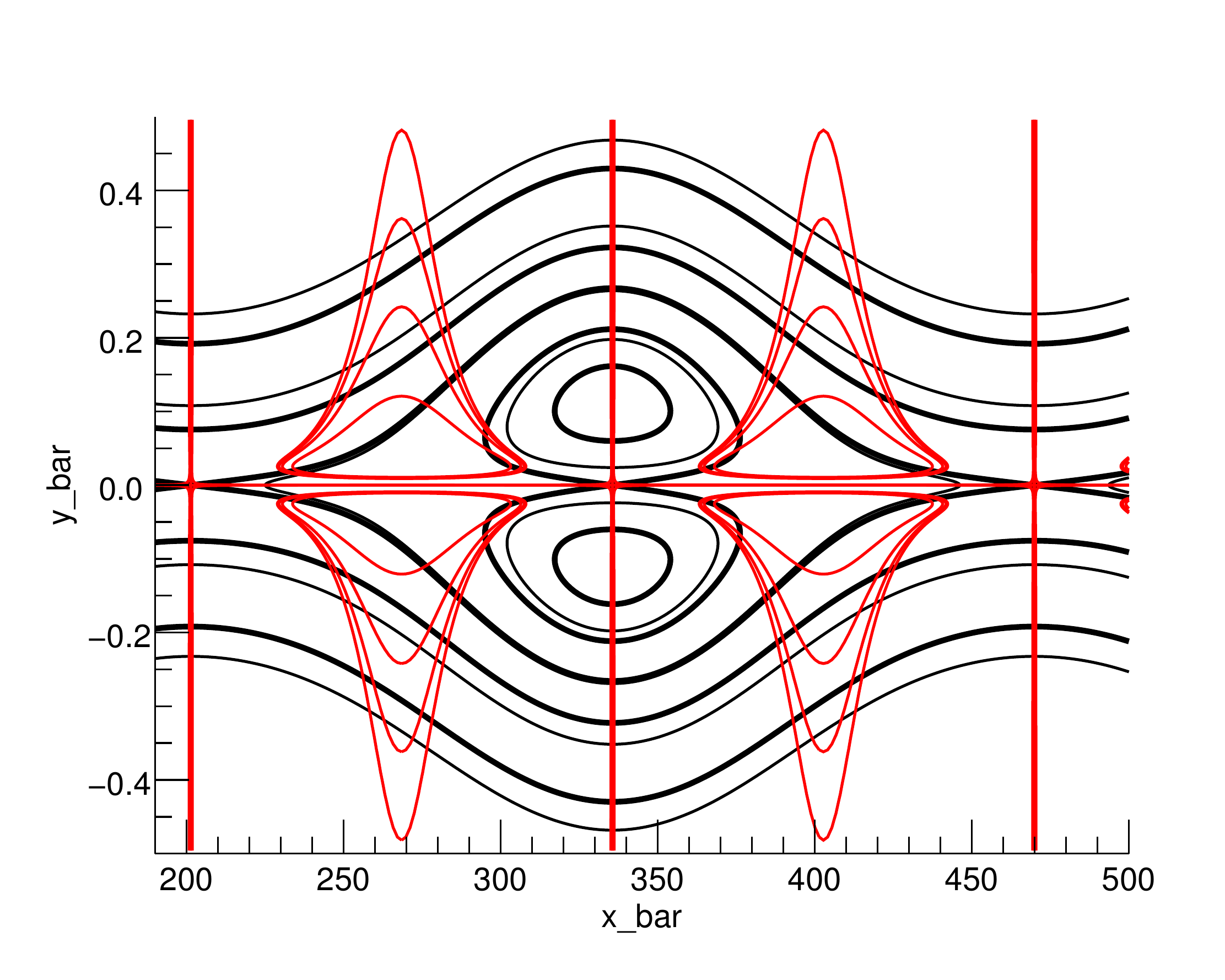}{0.3\textwidth}{(a)}
          \fig{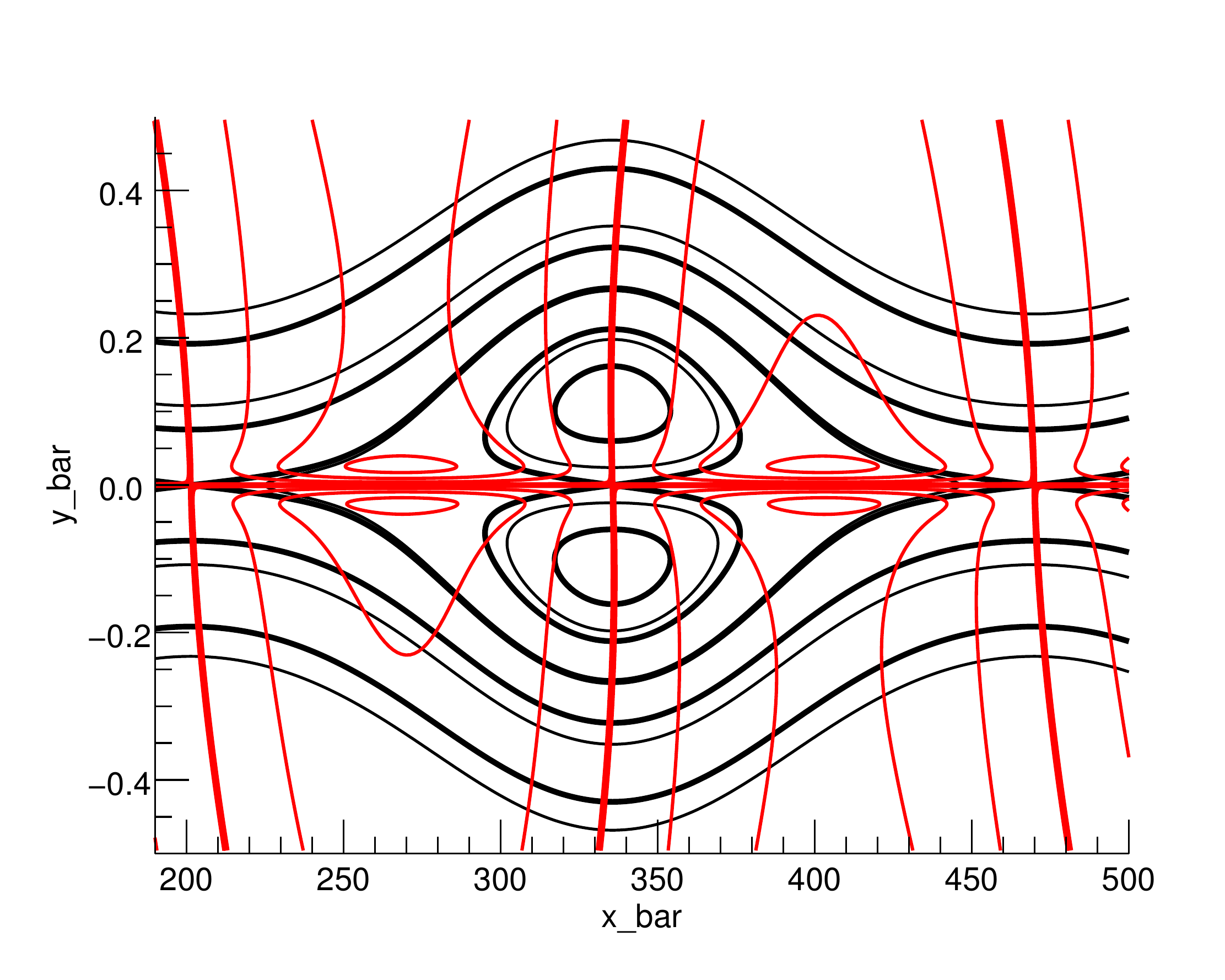}{0.3\textwidth}{(b)}
          \fig{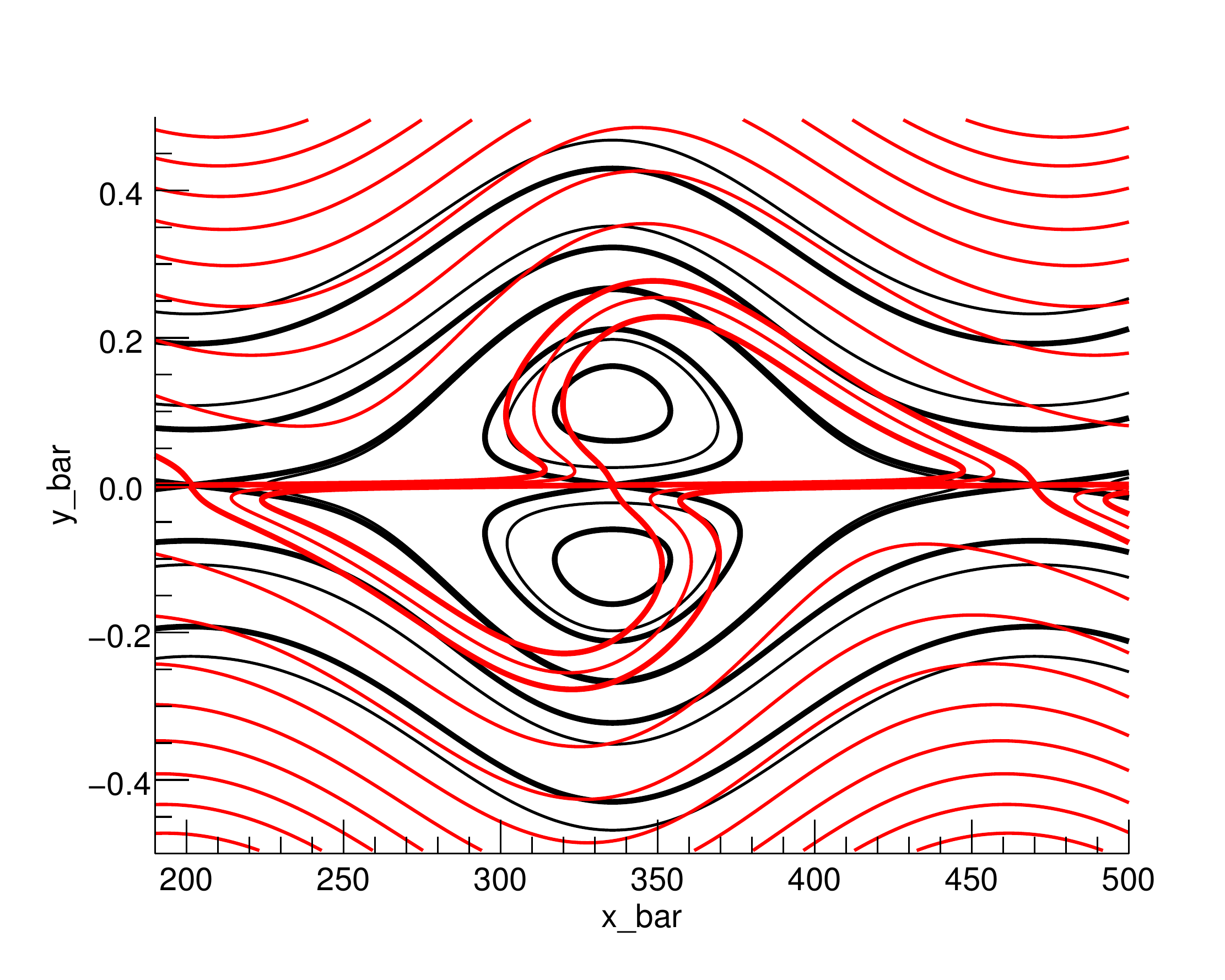}{0.3\textwidth}{(c)}
          }			
			
	\end{figure}

The full flux function and stream function consist of the equilibrium contributions added to these perturbations, times a small amplitude $\alpha$,
	
	\begin{equation}
	\label{eqn:A}
	A(x,y) = A_{0} + \alpha A_{1}~~,
	\end{equation}
	\begin{equation}
	\label{eqn:PHI}
	\Phi(x,y) = \Phi_{0} + \alpha \Phi_{1}~~.
	\end{equation}
These functions are contoured in Figure \ref{fig:contour} for $\alpha=0.0025$ and  three different values of $\bar{V}$: $\bar{V}=0$, $\bar{V}=1 \times 10^{-3}$, and $\bar{V} = 0.1$. \edit1{We choose these model parameters to illustrate the transition of the eigenmodes from zero shear flow to shear flow that is a significant fraction of the Alfv\'en speed.}	 $\Phi$ is contoured in red and $ A$ in black.  Magnetic field lines follow the black contours and velocity flow follows the red. Contour spacing is greatly exaggerated and chosen to highlight key features.   The magnetic structure in all three cases consists of large magnetic islands separated into two lobes each, with O-points at the lobe centers.  These double-lobed islands are separated from one another by 
X-points located at $\bar{k}\bar{x}= (4n+3)\pi/2$.  There is also an X-point at the center of each island, 
$\bar{k}\bar{x}= (4n+1)\pi/2$, separating the two lobes and their central O-points.  Remarkably, the flux functions are virtually indistinguishable in these three cases in spite of their very different shear flows.  This is because the imaginary part of 
$\bar{B}$ is extremely small compared to its real part, for all values of $\bar{V}$ we have considered.
	
	 With no background shear velocity (Figure \ref{fig:contour}a) flow is into the magnetic X-point on either end of the island along straight vertical paths and out along straight horizontal paths. The horizontal flows then converge on the X-points inside each island.  They exit these along vertical paths, filling the separated lobes composing each island.   This flow pattern brings new flux into the X-points, where it reconnects, and then moves into the island. Once in the island lines are reconnected a second time to form the two interior island lobes.  This flow pattern forms closed stream lines above and below the mid-plane of the current sheet.
	 
	 In Figure \ref{fig:contour}c Im$(\bar{u})_{max} \ll \bar{V} \ll v_a$. Shear flow of this speed is incapable of perturbing the magnetic field and is therefore forced to trace the perturbed magnetic field lines over most of eigenmode.  This behavior comes out of Equations (\ref{eqn:tm_in}) and (\ref{eqn:tm_momentum}) in the limit Im$(\bar{B}) \ll 1$ and $\bar{V} \gg {\rm Im}(\bar{u})_{\rm max}$ where Re$(\bar{u}) \rightarrow (G/F)~{\rm Re}(\bar{B})$.  
	 
	 Figure \ref{fig:contour}b shows an intermediate regime,  $\bar{V} \approx {\rm Im}(\bar{u})$, in which streamlines occur in two distinct classes. Far enough from the current sheet streamlines form open contours, following the perturbed field lines and not crossing the current sheet. Closer streamlines, on the other hand, are closed and circulate through magnetic X-points in distorted versions of the closed vortices in the shear-free case (\ref{fig:contour}a).  The distance from the current sheet where the streamlines transition from open to closed depends on the magnitude of the perturbation $\alpha$.

	The linear model of the TM/KH mode instability outlined above is not new. However, we hope that a more detailed view of the instability eigenmodes, for various values of background shear flow, help illuminate the nature of this combined instability.  Repeating the analysis has also allowed us to confirm our conjecture in the following section that the combined TM/KH instability can reproduce the behavior observed by \citet{Brannon2015}

\section{Modeling Observations}

\subsection{\edit2{Geometry}}
	During the flare observed by IRIS on April 14, 2014 a periodic sawtooth pattern was observed by \citet{Brannon2015} (\BLQ) to propagate with a phase velocity of 15 km s$^{-1}$ parallel to the ribbon and therefore perpendicular to the IRIS slit.  Line of sight (LOS) Doppler measurements of the fluid composing this sawtooth pattern found oscillations with an amplitude of $\approx \pm20$ km s$^{-1}$.  These velocities were found to be $\approx 180^{\circ}$ out of phase with the position on the slit. The observed behavior is best captured in Figure \ref{fig:Brannon11}.  It can also be seen in this figure that the amplitude of the oscillations along the slit are roughly 1 arcsecond. The Tearing/Kelvin-Helmholtz modes described above can produce features consistent with described observation. 
	
\begin{figure}	
	\caption{In the proposed magnetic field configuration a flow in the $\hat{x}$ direction, $\protect\overrightarrow{V}$ (blue), deflects overlying magnetic field, $\protect\overrightarrow{B}$ (black), at some angle $\theta$ to the inner recently reconnected magnetic field, $\protect\overrightarrow{B}$ (red).  This results in the labeled currents, $\protect\overrightarrow{J}$ (green).  The dashed box shows a blown up region of the secondary current sheet where we model the TM/KH Instability.  We also show the LOS of the observation to demonstrate how motion in the x-y plane translates to LOS Doppler velocities.}
	\label{fig:cartoon}
	
	\centerline{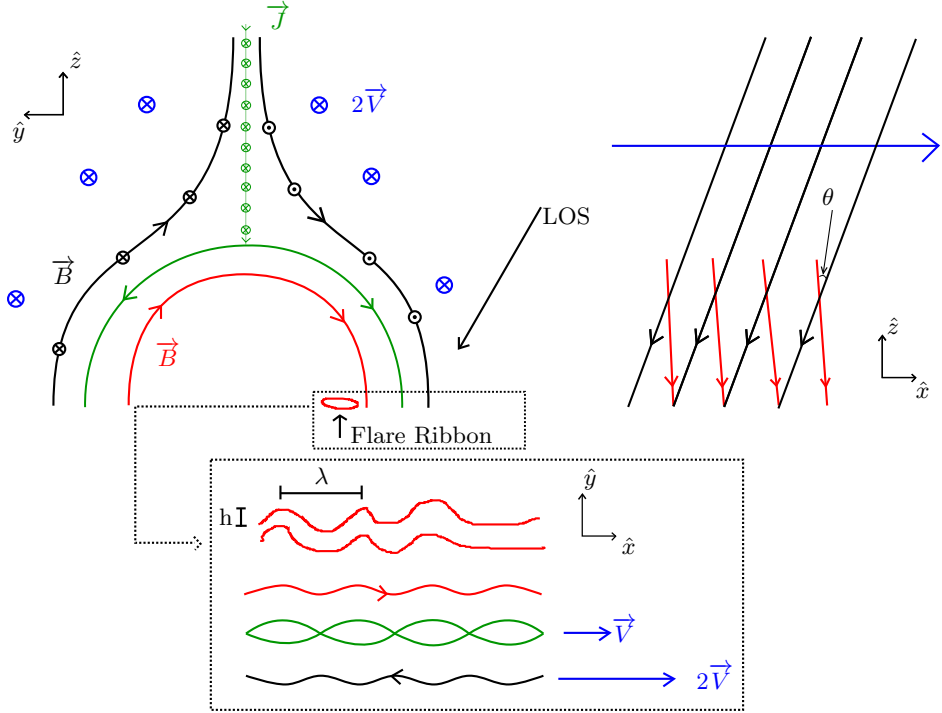} 

	\end{figure}

	We propose that a tearing mode had occurred within a secondary current sheet that had formed along the separatrix legs of the flare arcade.  A schematic version of the proposed geometry is shown in Fig. \ref{fig:cartoon}.  This differs from the geometry first put forth by BLQ, who expected the TM to occur within the main current sheet beneath the erupting flux rope.  We find that our geometry is able to better match all aspects of the observation.      
		
	The secondary current sheet separates newly reconnected arcade field lines from still-open flux connected to the flux rope above.  We propose that horizontal motion in this open flux, perhaps persisting from the eruption,  creates the the velocity shear as well the magnetic shear which is the current sheet itself.  The flux on the other side of this layer, consisting of newly closed arcade field lines, is expected to be stationary and undeflected.  These are the field lines in which energy has been deposited, so the ribbon itself should occur {\em within} this stationary flux, rather than at the secondary current sheet.
	
	The current in the secondary sheet flows {\em vertically}, so the tearing mode in it will produce a {\em horizontal} chain of islands, as illustrated in Fig. \ref{fig:cartoon}.  These will deform the magnetic field outside the sheet, and therefore the ribbon.  If the tearing mode propagates along the sheet, so will this deformation.   We propose that it is this propagating deformation which produces the moving sawtooth pattern in the ribbon.

	%We assume the instability has saturated with islands that have reached a maximum size proportional to $k^{-1}$ \citep{Biskamp1982} . We therefore set the parameter $\alpha = 1$ in Equations (\ref{eqn:A}) and (\ref{eqn:PHI}).
	
\subsection{\edit2{Introducing dimensions}}
	
	 \edit2{To help constrain parameters we assume that the linear instability has had sufficient time for $\sim10$ $e$-foldings during the typical flare time scale of $t\simeq100$ s.  We therefore set $\bar{\gamma} \bar{t} \propto 10$, to obtain the constraint} 
	
	\begin{equation}
	\label{eqn:gammamin}
	\bar{\gamma} \geq \frac{10\,a}{(100\, \mathrm{s})\, v_a}= \frac{\bar{k} \lambda}{2 \pi v_a\, (10\,\mathrm{s})}~~.
	\end{equation}
Here $\lambda$ is the wavelength of the sawtooth pattern from Figure \ref{fig:SJI} along the solar surface.  We have no direct measurement of the sheet's thickness, $a$, but we can use the observed wavelength, $\lambda$, to dimensionalize our expressions.  It is thus, in Eq.\ (\ref{eqn:gammamin}), that we use $\bar{k} = ka = 2\pi a/\lambda$ to formally eliminate $a$. 

	We assume that the horizontal motion of the open flux has deflected it by some small angle, say $10^{\circ}$, relative to the stationary, and undeflected arcade flux.  This angular difference, $\pm5^{\circ}$, produces the tangential discontinuity which is the secondary current sheet.  The field strength in the vicinity of the flare ribbon region was estimated by BLQ to be 150 G.  The single-sided  $10^{\circ}$ deflection will therefore produce a reconnecting component of $150\sin(5^{\circ})=13$ G.  Assuming a typical electron density of $3 \times 10^9\,{\rm cm^{-3}}$, gives the value $v_a = 500$ km s$^{-1}$, which was used in our non-dimensionalization.  
	
	The sawtooth pattern is observed to propagate with phase velocity perpendicular to the IRIS slit of $V' = 15$ km s$^{-1}$.
This was found in the plane-of-the sky (primed coordinates), but applied to motion horizontal motion some $40^{\circ}$ from disk center.  Foreshortening effects will reduce velocity $V$, in the horizontal $x$ direction, to an apparent velocity
	\begin{equation}
	\label{eqn:x_scale}
		V' = V\cos\theta_x~~.
	\end{equation}
where $\theta_x\simeq33^{\circ}$ is the horizontal distance from disk center.  The observed phase velocity of $15$ km s$^{-1}$ thus corresponds to a real horizontal velocity of $V \simeq 18.5 $ km s$^{-1}$.  Scaling this to the reconnection-component Alfv\'en speed, $v_a=500$ km s$^{-1}$ yields 
$\bar{V}=0.037$.  

\subsection{\edit2{The saturated state}}

\begin{figure}	
	\caption{Contours of TM/KH eigenfunctions for $S=\Lundquistmodel$, $\bar{k}=\kmodel$, and $V' = .037$  with $\alpha = 1$. The thick green line shows the trajectory of a particle that started at $[\bar{x},\bar{y}]=\posmodel$. }
	\label{fig:contour_sat}
	
	\centerline{\includegraphics[scale=.8]{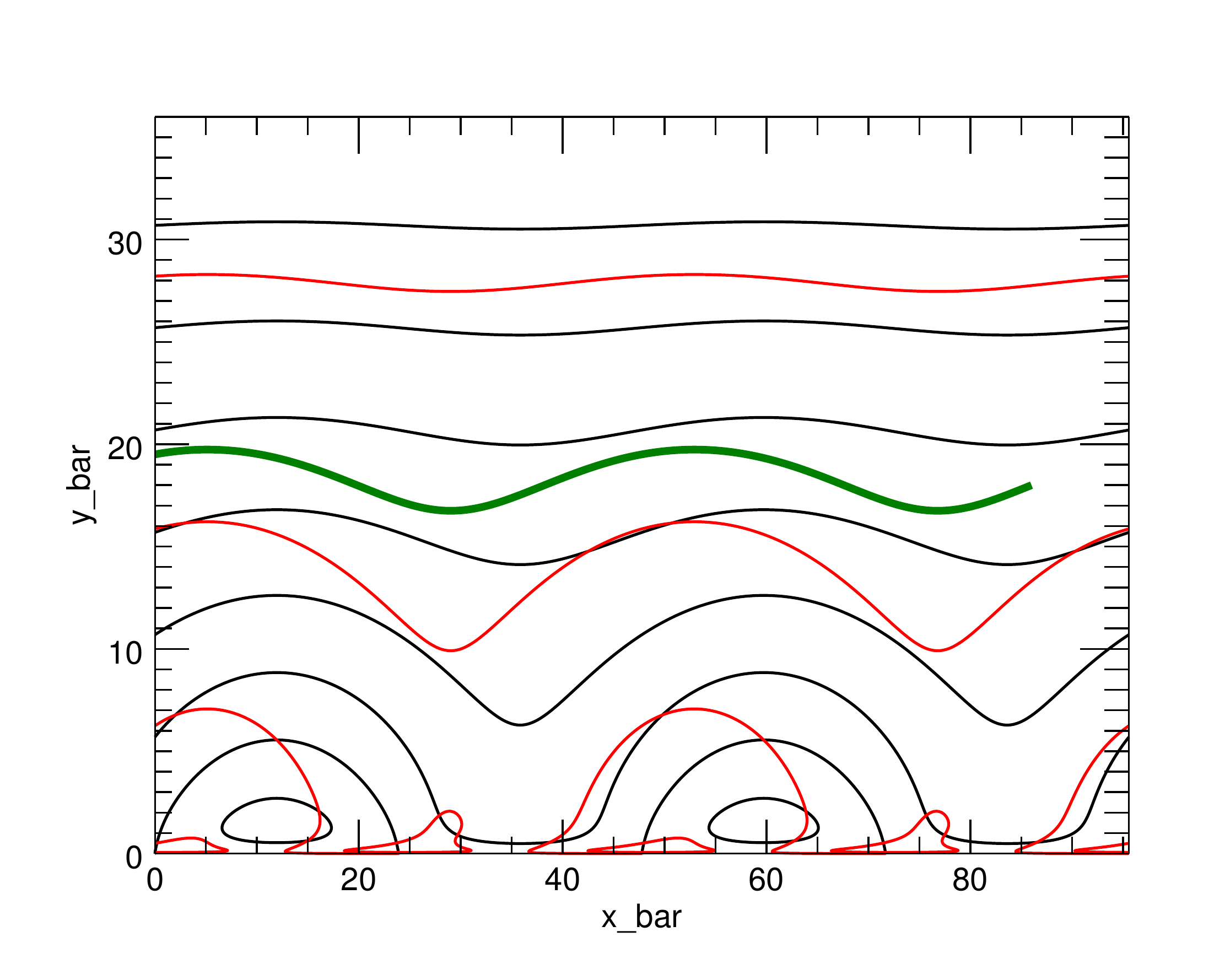}}

	\end{figure}

\edit2{We use these dimensions to construct a synthetic observation from the velocity structure of our model.  We do so first using the velocity structure given by the linear eigenmode, $\Phi_1(x,y)$, discussed above.  We construct the stream function by superposing this using the amplitude, $\alpha=1$, in eqs.\ (\ref{eqn:A}) and (\ref{eqn:PHI}).  Figure \ref{fig:contour_sat} shows magnetic field lines (black) and streamlines (red) of the resulting construction.  This specific eigenfunction results from $S=\Lundquistmodel$ and $\bar{k}=\kmodel$, which has a growth rate of $\bar{\gamma} = \gmodel$ in agreement with (\ref{eqn:gammamin}).  The green contour shows the trajectory of a fluid element that begins at $[x,y]=\posmodel$.}
	
\edit2{Because the growing linear eigenmode has been used in the non-linear regime, the stream lines (red) cross the field lines (black) in this construction.  The magnetic field evolution implied by continuing island growth requires this cross-field flow in the regions outside the diffusion region.   This cross-field flow will naturally cease when the linear phase ends and the island growth ends.}
	
\begin{figure}  
	
	\caption{Contoured eigenfunctions of an approximated saturated state. Here the instability growth has stopped and the reconnection inflow/outflow has gone to zero. The green contour shows the trajectory of a fluid element beginning at $[\bar{x},\bar{y}]=[86,13]$}
	\centerline{\includegraphics[scale=.8]{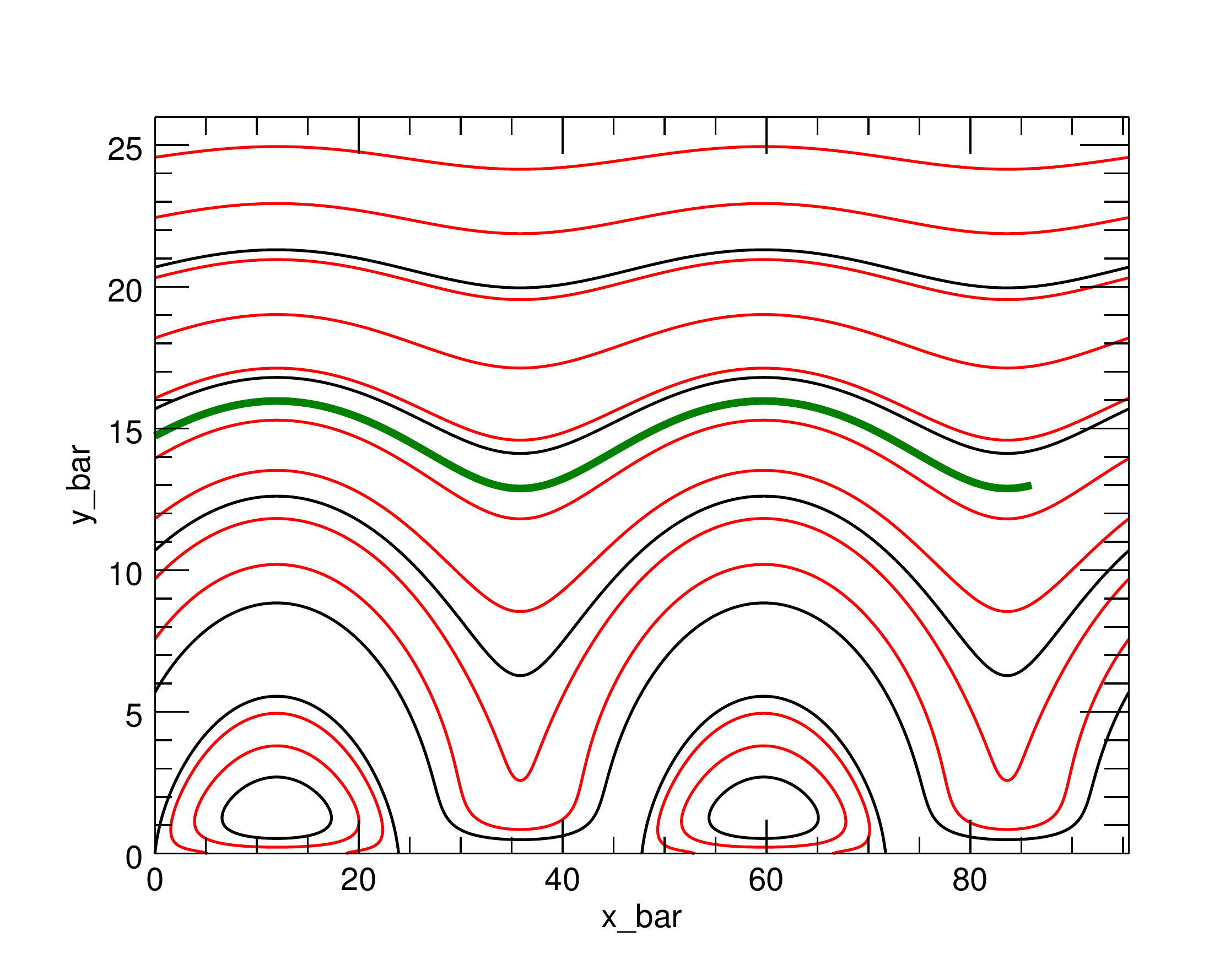}}
	\label{fig:saturated}
	\end{figure}

\edit2{Fully non-linear treatments of the tearing mode, in which island saturation occurs self-consistently, have found magnetic structures very similar to those produced by the linear modes with $\alpha=1$ \citep{Offman1993,Chen1997,Li2010,Zhang2011,Doss2015}.  However,these treatments find a velocity field very different from the linear eignmode.  In particular, the velocity must align with the magnetic field outside the diffusion region which is not the case with the unsaturated linear eigenmode, as seen in Fig.\ \ref{fig:contour_sat}.   Non-linear studies of island saturation in the presence of external shear have shown the saturated stream-function assumes, instead, a form $\Phi(x,y) \simeq (V_0/ B_0)A(x,y)$.  In this case the flow follows magnetic field lines exactly and matches the  external shear flow asymptotically.  Such a structure is clear in the non-linear simulation of \citet{Offman1993}. Using this instead of the linear eigenmode results in the stream lines (red) shown in Figure \ref{fig:saturated}.  A particular fluid element is traced in green, as it was in Figure \ref{fig:contour_sat}.  While the inter-relations of the magnetic field and stream functions differ significantly between these two constructions, it is noteworthy how similar the green flow lines appear.  In both cases the flow is made to oscillate by the magnetic field distortions produced by the islands} 

\subsection{\edit2{Transformation to Observed Quantities}}

\begin{figure}
	\caption{Panels a and b show the green trajectory from Figure \ref{fig:contour_sat} after undergoing the Galilean transform in Equation (\ref{eqn:lorentz}). Panels c and d show the same for the green trajectory in Figure \ref{fig:saturated}. All displacements have been foreshortened to account for off disk observation and scaled to physical units.   The green dot indicates the particles starting location.}
	\label{fig:traj}
	\centerline{\includegraphics[scale=.45]{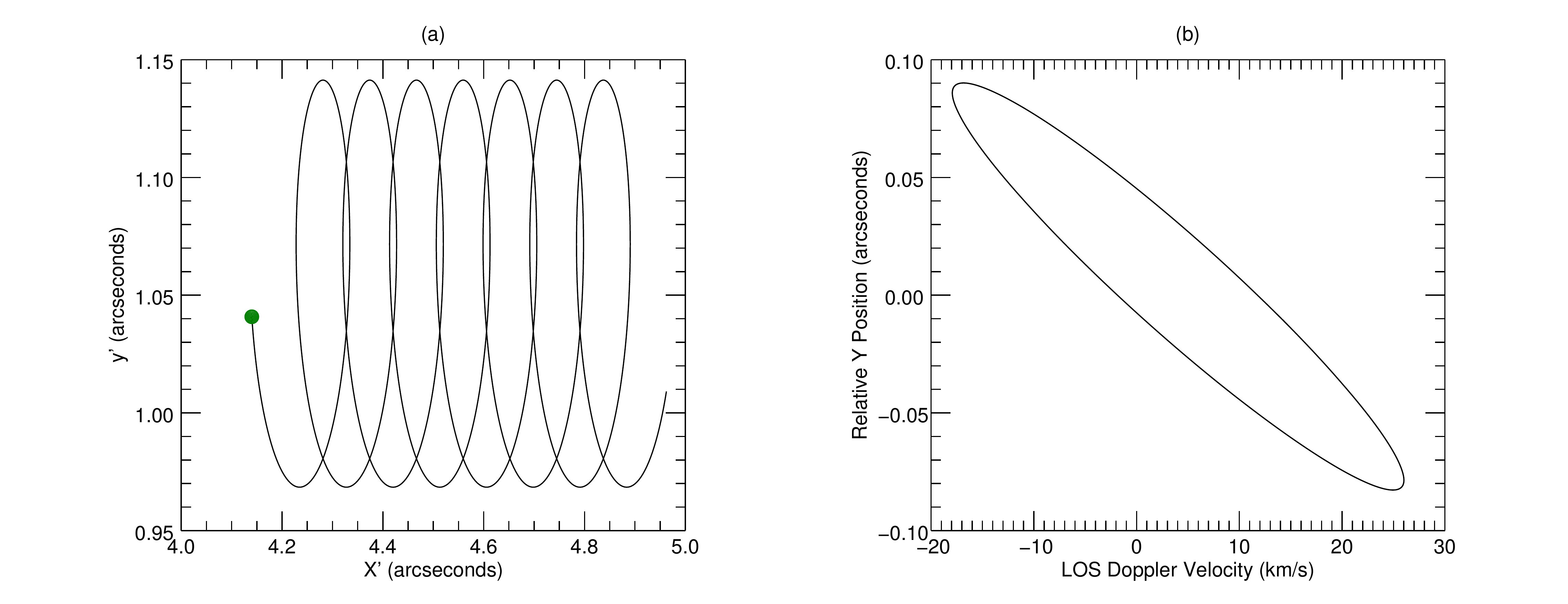}}
	\centerline{\includegraphics[scale=.45]{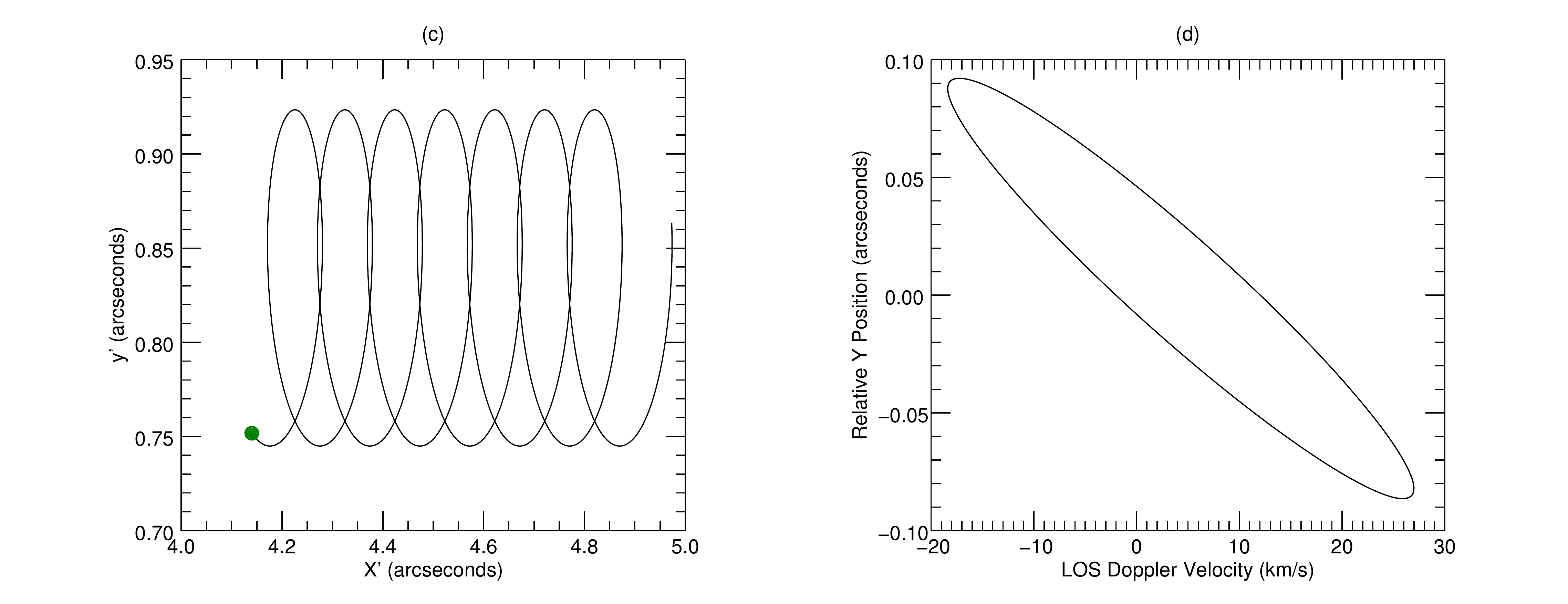}}
\end{figure}

	The analysis of the previous section was performed in a reference frame moving so as to make the velocity shear symmetric, thereby endowing the TM with zero phase velocity.  In our proposed model one side of the CS is stationary, while the other is moving horizontally at some speed $v_0$ (see Fig. \ref{fig:SJI} and \ref{fig:cartoon}).  This means the analysis references frame is translating at $V=v_0/2=18.5$ km s$^{-1}$ to produce the phase velocity observed in the flare ribbon. Solutions for an asymmetric shear flow are found through a Galilean transform of the horizontal dimension,
	\begin{equation}
		\label{eqn:lorentz}
		\bar{X} = \bar{x} - \bar{V} \,\bar{t}~~.
	\end{equation}
Note again that $\bar{V}$ is not the observed speed, but the speed of propagation along the solar surface. The transform is such that flow below the current sheet is $2 \bar{V} = .074$ and zero above just as in Figure \ref{fig:SJI}. In this reference frame an observer would see the train of islands propagating in the direction of the non-zero shear flow, positive $\bar{X}$, with a velocity equal to $\bar{V}$. A fluid element following the green \edit2{trajectories in either Figures \ref{fig:contour_sat} or \ref{fig:saturated}, loses most of its net velocity along the current sheet.  Instead it undergoes elliptical motion, Figures \ref{fig:traj}a and Figures \ref{fig:traj}c respectively, as the the islands move past.} % We also must account for the projection effects of the observation being off disk center.  

\edit2{To construct the flows in Figure \ref{fig:traj} }we move from our non-dimensional model to physical units and observed quantities in the \textit{Plane-Of-Sky} (POS) frame. Model quantities are foreshortened through the following transformations to observed quantities (denoted by primes):  
	
	\begin{equation}
		y' = \bar{y}\cos(\theta_y)~~, 
	\end{equation}
	\begin{equation}
		X' = \bar{X}\cos(\theta_x)~~,
	\end{equation}
\edit2{where viewing angles are
	\begin{equation}
		\theta_x = \sin^{-1}(x_{\mathrm{POS}}/R_{\odot})~=~33^{\circ}~~,
		\quad \theta_y = \sin^{-1}(y_{\mathrm{POS}}/R_{\odot})~=~ -11^{\circ}~~,
        \end{equation}
for our region located at $x_{\mathrm{POS}}=553''$, $y_{\mathrm{POS}} = -188''$, when the apparent semi-diameter is $R_{\odot} = 956''$.}

Spatial dimensions are scaled to arcseconds through current sheet thickness $a$.  For each run of the model we specify $\bar{k}$, which with a fixed $\lambda$ also specifies the current sheet thickness, $a$. This is done in the following way:
	
	\begin{equation}
		\bar{k} = ka = \frac{2\pi a}{\lambda}~~, \quad \lambda = \frac{\lambda'}{\cos(\theta_x)}~~,~\mathrm{and}~\lambda'=2.3''~~.
	\end{equation}
In this \replaced{titled}{tilted} frame a portion of the motion in both the $X$ and $y$ dimension are along the IRIS line of sight.
	\begin{equation}
		V_{\mathrm{LOS}} = \sin(\theta_x)\frac{d\bar{X}}{dt} + \sin(\theta_y)\frac{d\bar{y}}{dt}~~.
	\end{equation}
This velocity is scaled to physical units through $v_a = 500$ km s$^{-1}$. These transformations and scalings allow for a direct comparison \replaced{of}{between} Figure \ref{fig:Brannon11} and \ref{fig:traj}.  

\edit2{Figures \ref{fig:traj}b and \ref{fig:traj}d} show our model's reproduction of Figure \ref{fig:Brannon11} using the parameters above.  \edit2{These use the fluid trajectories from Figure \ref{fig:contour_sat} or \ref{fig:saturated} respectively, and obtain very similar results for reasons already mentioned.  For both constructions we find that we can produce similar results using a wide range of other parameters.}  

It is worth noting that in \edit2{Figures \ref{fig:traj}b and \ref{fig:traj}d, as well as in the observations of Figure \ref{fig:Brannon11}, line-of-sight} velocities oscillate from roughly $-15$ km s$^{-1}$ to $+25$ km s$^{-1}$.  The DC offsets in the model \edit2{versions} come from the fact that oscillations are not perfectly elliptical as is seen in \edit2{Figures \ref{fig:traj}a and \ref{fig:traj}c}.  The observational data is probably consistent with this behavior within the measurement uncertainties. It is also worth noting that the Doppler velocity is not perfectly 180$^{\circ}$ out of phase with relative $y$ position.  This is due to the fact that motion in the $\bar{y}$ direction has a component of velocity along the LOS in the tilted reference frame of the observation. The measured amplitude of the oscillation along the slit is harder to reproduce.  This we can attribute to the fact that our linear model is incapable of generating islands that are truly saturated in height.  

The measured amplitude of the sawtooth pattern is roughly 1 arcsecond which is larger than $\lambda /2$.  Therefore, our linear model would never be able to exactly match observation.  Regardless, we are able to report that our simple model gets us within a factor of 5 from true measured values.  
	
	The amplitude of both position and velocity depend on how far the fluid element is from the sheet, which is the starting value of $\bar{y}$.  Elliptic motion in the solar frame are produced from streamlines which are {\em open} in the analysis frame.  Closed streamlines will produce motion moving with the analysis frame. Moving closer to the current sheet gives higher amplitudes in position and velocity but also deviates from the mostly elliptical trajectory. Regardless, the elliptical motion and resulting phase portrait of Figure \ref{fig:traj} are achievable for a variety of initial conditions. Doppler velocities on order 20 km s$^{-1}$ and elliptical trajectories with sufficient amplitude have been produced for $S = 10^4$ to $10^7$ and $\bar{k} = .01$ to $.5$.  The most limiting factor, and the reason we chose $S = \Lundquistmodel$ for producing Figures \ref{fig:contour_sat} -- \ref{fig:traj}, was getting a growth rate high enough to satisfy Equation (\ref{eqn:gammamin}).

\section{Discussions and Conclusions}
We have used a simple model of the tearing mode instability, defined in Section 2, to provide evidence that a tearing mode with asymmetric shear flow could produce the flare ribbon oscillations observed by \citet{Brannon2015}. In our model asymmetric shear flow provides a net phase velocity to the instability, a key element of the findings. It also produces elliptical horizontal motions of fluid elements of the kind inferred by BLQ on the basis of the phase relationship between oscillations in velocity and position.  We propose a scenario in which a secondary current sheet develops at the separatix of the flaring field.  A tearing mode in that secondary current sheet produces fluid motions similar to those observed.  We rely on the measurement of both the sawtooth pattern wavelength, $\lambda' =2.3"$, and  phase velocity of 15 km s$^{-1}$ as well as a typical flare timescale of 100 s to re-dimensionalize our model.  Under these constraints we are able to reproduce the observed behavior of Figure \ref{fig:Brannon11} in \edit2{Figure \ref{fig:traj}b and \ref{fig:traj}d} to a satisfactory level of accuracy.

\edit1{Our model is constrained by observations of the parameters} $\lambda$, $\Delta y$, $\Delta v_{\rm LOS}$, and a phase velocity of 15 km s$^{-1}$ as well as achieving a minimum growth rate to satisfy Equation (\ref{eqn:gammamin}). These constraints limit our solution to a small area of parameter space. Assuming that $\bar{k} = \bar{k}_{\rm max}$ the only free parameters are $S$ and the distance $y$ from the CS. We can reasoanbly fit the observations for $S= 1-1.25 \times 10^4$, or subsequently $\bar{k} = 1.22-\kmodel$. This range of $\bar{k}$, along with a measured wavelength $\lambda = 2.3"$, yields a current sheet thickness $a \simeq 39 - 42$ km. This is considerably smaller than current sheets reported through other observations. \citet{Savage2010} observe a long, thin, bright structure above an arcade with a thickness of several times $10^3$ km. Others have measured current sheet thicknesses of similar magnitude \citep{Seaton2016}. Models have shown that these large observed thicknesses may be the result of a layer of hot dense plasma surrounding the current sheet rather than the current sheet itself \citep{Seaton2009,Reeves2010,Yokoyama1998}. This leads us to believe that a current sheet on order $10$ km is a realistic estimate for the April 18 flare.

Resistivity sets a lower limit on the CS thickness given by the steady Sweet-Parker model: $a=LS^{-1/2}$, where $L$ is the length of the CS.  The entire flare ribbon, in this case, has a length $L\sim 50$ Mm.  Classical Spitzer resistivity, yields $S\sim 10^{12}$ from which we would predict a lower limit of $a=50$ m.  It is possible that the resistivity in the vicinity of the CS is enhanced, perhaps by turbulence or instability \citep{Strauss1986} to a level where $S\sim 10^6$ for which a Sweet-Parker CS would have the thickness we infer.  It is also possible that the sheet has not fully collapsed, and its thickness is instead set by the dynamics of its formation.  It is not, after all, the flare CS, and is only formed as a secondary consequence of the flare.

The geometrical scenario proposed in Figure \ref{fig:cartoon}, necessary for this observation, \replaced{do}{does} not follow from the standard CSHKP model, and \replaced{are}{is} not inevitable in a two-ribbon flare.  The lack of sawtooth reports from the many other flare ribbon observations made by IRIS suggests that sheared, tearing-unstable secondary current sheets, are not a common to all flares.   It would seem they have occurred at least twice -- in the flares of 2014-04-18 and again on 2014-09-10.   A survey of several other, seemingly suitable IRIS ribbon observations has shown that sawteeth are indeed uncommon, but other candidates seem to have occurred \citep{Roegge2017}.  Shear flows responsible for deflecting the overlying magnetic field must be fast enough and sustained long enough for a secondary current sheet to develop and become unstable. Furthermore, if the wavenumber or the amplitude were smaller than that of April 18, the pattern may be too small to resolve with IRIS.  Indeed, the pattern pattern observed in that flare is marginally detectable in coincident AIA images (BLQ). 

%A propagating flare ribbon, despite being compelling evidence of a TM instability occurring on the Sun, does little to unveil the reason behind solar flares and the mechanism of their energy release.  

The proposed shear flow, embedding the CS, is essential for explaining the observed phase velocity of the sawtooth.  It is plausible that the eruption, initially traveling at many hundreds of km/s, left a slower, but  persistent, flow in its wake.  We have not, however, seen any clear evidence for this proposed flow, aside from the propagation of the sawtooth itself, in any of the data we examined.  It is possible that the shear flow and the tearing mode occur higher up the CS, away from the chromospheric ribbon.  The line-tying effect of the lower boundary could suppress the flow, while still permitting the form of the instability to be imprinted on the chromosphere. A more complicated three-dimensional geometry of this kind would need to be explored in a model more sophisticated than our two-dimensional version of the standard TM model.  Our model scenario calls for the horizontal shear flow just discussed, as well as a current sheet flowing vertically along the separatrix -- this is the secondary current sheet.  Such a secondary current sheet was reported by \citep{Janvier2014} in vector magnetograms data from a much larger flare.  In spite of the favorable orientation of that flare, measurement of this weak secondary current sheet proved extremely difficult.  It is nevertheless possible that a similar careful analysis of SDO/HMI data could reveal the presence  of the secondary current sheet in the 18 April, 2014 flare.

\edit2{We have used a simple two-dimensional, analytic model of a tearing mode in order to demonstrate, as simply as possible, the viability of our explanation of the observation.  An analytic model allows us to explore parameter space and to easily re-scale the results for comparison to observations.  The simplified  analytic model is, however, subject to limitations.  We are unable to self-consistently describe non-linear phases including saturation.  Instead we constructed velocities fields intended to represent two extreme limits.  Figure \ref{fig:contour_sat} shows an unsaturated state produced from the velocity field of the growing linear eigenmode.   Figure \ref{fig:saturated} approximates the fully saturated state, where flow must follow the magnetic field.  Remarkably, both these approximate, limiting cases yield observational signatures, compared side by side in Figure \ref{fig:traj}, which are virtually indistinguishable from one another.  This is no coincidence.  The basic feature revealed by the observation is flow near the current sheet being modulated by a moving chain of magnetic islands.  This produces elliptical motion of the kind reported by \BLQ, and our model implements the explanation original given in that work.  A self-consistent non-linear model is beyond the scope of our exploratory study, but we believe it would produce velocity structures falling between the two extreme approximations we do consider.  Observations synthesized from it would therefore probably resemble the indistinguishable versions produced by the extreme cases.  It would therefore match the observations for the same reasons they do.}

If our hypothesis \edit2{stands up under} further scrutiny, the April 18 observation provides novel evidence for a TM in the solar corona. Plasmoid observation, and the presence of a TM instablilty, provide insight into the onset of fast magnetic reconnection.  It is also noteworthy that while the TM occurs during a flare, it does not occur in the flare's primary current sheet.  The sawtooth would thus seem to be a secondary effect offering only limited insight into the flare itself.

\acknowledgments 
The authors would like the thank Dr. Sean Brannon for his work with this observation. This work has been supported by NASA grants NNX14AD43G and NNX16AH04G as well as contract 8100002702 from Lockheed Martin to Montana State University.  IRIS is a NASA small explorer mission developed and operated by LMSAL with mission operations executed at NASA Ames Research center and major contributions to downlink communications funded by ESA and the Norwegian Space Centre.

\end{document}